\documentclass[12pt]{article}
\usepackage{geometry}
\usepackage{amsmath}
\usepackage{amsfonts}          
\usepackage{amssymb}
\usepackage{xspace} 
\usepackage{graphicx}   

\def\appendix#1{
  \addtocounter{section}{1}
  \setcounter{equation}{0}
  \renewcommand{\thesection}{\Alph{section}}
 \section*{Appendix \thesection\protect\indent \parbox[t]{11.715cm} {#1}}
  \addcontentsline{toc}{section}{Appendix \thesection\ \ \ #1}
  }

\renewcommand{\thefootnote}{\fnsymbol{footnote}}

\numberwithin{equation}{section}

\newcommand{\be}{\begin{equation}}
\newcommand{\ee}{\end{equation}}
\newcommand{\ba}{\begin{aligned}}
\newcommand{\ea}{\end{aligned}}
\newcommand{\half}{{1\over 2}}
\newcommand{\ie}{{\it i.e.}}

\newcommand{\p}{\partial} 
\newcommand{\Vect}{{\mathrm{Vect}}}

\def\Ann {{\rm{Ann}}}

\makeatletter
\def\sla@#1#2#3#4#5{{%
  \setbox\z@\hbox{$\m@th#4#5$}%
  \setbox\tw@\hbox{$\m@th#4#1$}%
  \dimen4\wd\ifdim\wd\z@<\wd\tw@\tw@\else\z@\fi
  \dimen@\ht\tw@
  \advance\dimen@-\dp\tw@
  \advance\dimen@-\ht\z@
  \advance\dimen@\dp\z@
  \divide\dimen@\tw@
  \advance\dimen@-#3\ht\tw@
  \advance\dimen@-#3\dp\tw@
  \dimen@ii#2\wd\z@
  \raise-\dimen@\hbox to\dimen4{%
    \hss\kern\dimen@ii\box\tw@\kern-\dimen@ii\hss}%
  \llap{\hbox to\dimen4{\hss\box\z@\hss}}}}
\def\slashed#1{%
  \expandafter\ifx\csname sla@\string#1\endcsname\relax
    {\mathpalette{\sla@/00}{#1}}%
  \else
    \csname sla@\string#1\endcsname
  \fi}
\makeatother

\begin{document}


\thispagestyle{empty}
\begin{flushright}\footnotesize
\texttt{hep-th/0609084}\\
\texttt{CALT-68-2610}\\
\texttt{DESY 06-155}\\
\texttt{ZMP-HH/06-015}\\
\vspace{2.1cm}
\end{flushright}

\renewcommand{\thefootnote}{\fnsymbol{footnote}}
\setcounter{footnote}{0}
\begin{center}
{\Large\textbf{\mathversion{bold} 
 T-duality with $H$-flux:\\
 non-commutativity, T-folds  and $G\times G$ structure
}\par}

\vspace{2.1cm}

\textrm{Pascal Grange$\,^\ast$ and Sakura Sch\"afer-Nameki$\,^{\ast, \sharp}$}

\vspace{1cm}

\textit{$^\ast$ II. Institut f\"ur Theoretische Physik der Universit\"at Hamburg\\
Luruper Chaussee 149, 22761 Hamburg, Germany} \\

\vspace{4mm}

\textit{$^\ast$ Zentrum f\"ur Mathematische Physik, Universit\"at Hamburg\\
Bundesstrasse 55, 20146 Hamburg, Germany} 

\vspace{4mm}

\textit{$^\sharp$ California Institute of Technology\\
1200 E California Blvd., Pasadena, CA 91125, USA
}

\vspace{6mm}
\texttt{pascal.grange@desy.de, ss299@theory.caltech.edu}

\vspace{3mm}


\par\vspace{1cm}

\textbf{Abstract}
\end{center}

\noindent
Various approaches to T-duality with NSNS three-form flux are reconciled. 
Non-commutative torus fibrations are shown to be the open-string version 
of T-folds. The non-geometric T-dual of a three-torus with uniform flux is 
embedded into a generalized complex six-torus, and the non-geometry is 
probed by D0-branes regarded as generalized complex submanifolds. The 
non-commutativity scale, which is present in these compactifications, is 
given by a holomorphic Poisson bivector that also encodes the variation of 
the dimension of the world-volume of D-branes under monodromy. This 
bivector is shown to exist in $SU(3)\times SU(3)$ structure 
compactifications, which have been proposed as mirrors to NSNS-flux 
backgrounds. The two $SU(3)$-invariant spinors are generically not 
parallel, thereby giving rise to a non-trivial Poisson bivector. 
Furthermore we show that for non-geometric T-duals, the Poisson bivector 
may not be decomposable into the tensor product of vectors.

\vspace{5mm}

\vspace*{\fill}

\newpage
\setcounter{page}{1}
\renewcommand{\thefootnote}{\arabic{footnote}}
\setcounter{footnote}{0}


\tableofcontents


\section{Introduction}

Compactifications with $H$-flux are known to give rise 
to topology changes and even to non-geometric situations
when T-duality is performed along directions which have non-trivial support of the NSNS $H$-flux \cite{Alvarez:1993qi, Gurrieri:2002wz, Kachru:2002sk, Hellerman:2002ax, Fidanza:2003zi, Bouwknegt:2003vb,Mathai:2004qc, Hull:2004in}. 
Non-geometry occurs for example in the
very simple situation of a three-torus endowed with an $H$-flux
proportional to its volume form. Consider namely the three-torus as a
trivial $T^2$-fibration over a circle.  Upon T-duality along the
fibre, the metric picks up a factor that makes it shrink under
monodromy around the base circle. The monodromy around the base is 
a non-trivial element of the $O(2,2;{\mathbb{Z}})$ group acting
on the two-torus. This prevents a three-dimensional global Riemannian
description from existing. Further T-dualizing along the base leads to
more pathological situations, where points do not exist even in a
local coordinate patch, and the fibres are conjectured to become
non-associative
\cite{Bouwknegt:2004ap, Shelton:2005cf}. We will restrict ourselves to the case of
two T-dualities, and assume that local coordinate patches do exist.
Progress in the description of non-associative T-duals was achieved in
the recent paper \cite{Ellwood:2006my}, which also contains
observations on the open-string metric and non-commutativity for two
T-dualities that have some overlap with ours.

Essentially three conjectures have been put forward for the
description of the T-dual of a torus with $H$-flux: 
\begin{itemize}
 \item[{\bf (I)}] Field of non-commutative tori: Mathai and Rosenberg
  proposed that T-dualizing along a two-torus with non-zero $H$-flux
  yields a fibration by (or more precisely: field of) {\it{non-commutative}} tori.
  In particular, this fibration is encoded in a closed one-form, which is obtained by integrating the 
  NSNS flux along the fibre directions  \cite{Mathai:2004qc, Mathai:2004qq, Mathai:2005fd}.
  \item[{\bf (II)}]  T-folds: 
   these are spaces where T-dualities can act as transition
   functions between local patches \cite{Hull:2004in}.
   The T-dualized directions are
   doubled, and T-duality transformations may patch the doubled fibres
   together. A sigma model with a T-fold as its target space was
   proposed, and its boundary conditions were studied in 
   \cite{Dabholkar:2005ve, Hull:2006va, Hull:2006qs, Lawrence:2006ma,Hellerman:2006tx}.
 \item[{\bf (III)}]  $G\times G$ structure compactifications: 
    $SU(3)\times SU(3)$ structure manifolds are characterized in terms of a pair of
    pure spinors, constructed as bilinear combinations of a pair
    $SU(3)$-invariant spinors of ${\mathrm{Cliff}}(6)$. In case the
    $SU(3)$-invariant spinors are not parallel to each other, their
    linear independence is encoded by a non-vanishing one-form, and
    the discrepancy between left- and right-moving complex structures
    is a potential source of non-geometry and/or
    non-commutativity. Moreover, \cite{GLW, Benmachiche:2006df}
    suggest the relevance
    of $SU(3)\times SU(3)$ structures for mirrors of NSNS flux compactifications.
  \end{itemize}

 These directions of research have developed somewhat independently
 from each other, and it is natural to ask if they are compatible.  It
 is also natural to expect that techniques from generalized complex
 geometry \`a la Hitchin and Gualtieri
 \cite{Hitchin:2004ut,Gualtieri:2003dx} should
 bring some insights into the problem for at least two reasons:

 firstly, generalized complex (GC) spaces have been related to
 non-commutativity in two instances: a non-commutativity scale is
 induced by the $(0,2)$ component of a $B$-field \cite{Kapustin:2003sg},
 and the master equation of the generalized B-model
 \cite{Pestun:2006rj} admits deformations by holomorphic Poisson
 bivectors into a Poisson sigma model, which is known to induce
 star-products in the algebra of observables \cite{Cattaneo:1999fm}; 

 secondly, the doubling of the torus fibres in T-folds reminds one of the sum
 of tangent and cotangent spaces considered in generalized complex
 geometry. But GC spaces have more structure than T-folds, indeed, in
 \cite{Hull:2004in,Lawrence:2006ma} T-folds were pointed out to be a real
 version of GC spaces. Moreover, elements of $O(2,2;{\mathbb{Z}})$
 called $B$-transforms and $\beta$-transforms act on maximally
 isotropic subspaces as symmetries of the inner product.\\
 
We shall therefore use as a main technical tool the geometry of pure
 spinors, that are in one-to-one correspondence with generalized
 complex branes, and building blocks for $SU(3)\times SU(3)$ structure
 compactifications.

Our conjectures, which we will justify in the case of tori with $H$-flux, are:
\begin{itemize}
 \item(I) vs. (II):
    The proposal (I) by Mathai and Rosenberg claims that the T-dual to a $T^3$ compactification with 
    $H$-flux along two of the T-dualized directions yields a non-commutative torus fibration. This is 
    reconciled with Hull's T-fold proposal by showing that the metric seen by the open strings on a T-fold 
    is precisely the one on the non-commutative torus fibration. Thus, the proposal (I) is the open-string 
    version of (II).
    This connection is discussed from various independent angles in sections 2, 3 and 4. 
 \item (II) vs. (III): 
    when both approaches are applicable as for the $T^6$ with $H$-flux, 
    they yield the same T-dual or mirror geometry. 
 \item(III) vs. (I): 
    We show that for a generic $SU(3) \times SU(3)$ structure compactification, 
    where the two $SU(3)$-invariant spinors are not aligned, there exists a Poisson bivector which parametrizes 
    non-commutative deformations. The non-commutativity is however again only relevant for the 
    open-string sector. This relation is discussed in section 5. As for the mirror of a six-torus 
    with $H$-flux, we observe that the Poisson bivector can in fact not be decomposed in terms of vectors, 
    which seems to indicate that not all the possible non-commutativity scales are inherited from 
    $SU(3) \times SU(3)$ structures. 
\end{itemize}


\section{T-folds and non-commutative tori}

In this section we shall mainly be concerned with the connection between non-commutativity and T-folds. 
We shall study this in the case of the simplest non-trivial example, which already illustrates the main point: the Mathai--Rosenberg 
non-commutative torus-fibrations are the open-string version of T-folds. This observation will then be discussed from the generalized geometry point of view in the next section.

The simplest example that exhibits all the key features is the $T^3$-compactification with $k$ units of NSNS three-form flux $H\in H^3(T^3, \mathbb{Z})$. We shall generally refer to NSNS-flux supported on a torus bundle $E$ with base $B$ and fibre $F$ of the type 
$H\in H^n(F) \otimes H^{3-n}(B)$ as an $n$-legged $H$-flux. Thus, the one-legged case is known to have a purely geometric T-dual. Our main focus is on the two-legged case, which will be shown to have a non-geometric T-dual. 

In order to understand the T-dual along two fibre directions, we consider the three-torus as a $T^2$-bundle over $S^1$ (parametrized by $x$) and dualize along the fibre directions parametrized by $y$ and $z$. The 
metric and $B$-field can be chosen as
\be
  ds^2 = dx^2 + dy^2 + dz^2\,,\qquad
  B    =  k x \, dy \wedge dz \,.
\ee
Due to the  $B$-field the monodromy $\mathcal{M}_k$  around the $S^1$ is non-trivial
 and reads \be
 \mathcal{M}_k = 
 \left(
    \begin{array}{cccc}
    1 & 0 & 0 & 0 \cr
    0 & 1 & 0 & 0 \cr 
    0 & -k& 1 & 0 \cr
    k & 0 & 0 & 1 
    \end{array}
 \right)\,,
\ee
in a basis adapted to the coordinates $(y,z,\tilde{y},\tilde{z})$, where $\tilde{y}$ and $\tilde{z}$ are T-dual to $y$ and $z$.
Naively applying the standard Buscher rules along the fibres yields the T-dual background
\be\label{TFoldData}
 ds^2 = dx^2 + {1\over 1+ k^2 x^2} (dy^2 + dz^2) \,, \qquad
 B    = {k x \over 1+ k^2 x^2} \, dy \wedge dz \,,
\ee
and the monodromy obtained after action of the T-duality matrix 
\[g_{yz}= \left( \begin{array}{cccc}
0 & 0 & 1 & 0 \\
0 & 0 & 0 & 1 \\
1 & 0 & 0 & 0 \\
0 & 1 & 0 & 0\\
 \end{array} \right)\]
 along the fibres is
\be
 \mathcal{W}_k = g_{yz}^{-1} \mathcal{M}_k g_{yz} =
 \left(
    \begin{array}{cccc}
    1 & 0 & 0 & -k \cr
    0 & 1 & k & 0 \cr
    0 & 0 & 1 & 0 \cr
    0 & 0 & 0 & 1 
    \end{array}
 \right).
\ee
As this is a non-trivial element (which is not merely a $B$-field shift or an element in 
the geometrically acting 
$SL_2 (\mathbb{Z}) \times SL_2(\mathbb{Z})$) of the T-duality group 
$O(2,2;\mathbb{Z})$, the resulting space is an example of a T-fold as defined by Hull in \cite{Hull:2004in}.

The alternative proposal by Mathai and Rosenberg \cite{Mathai:2004qc, Mathai:2004qq, Mathai:2005fd} claims that the T-dual is a field $C$
of non-commutative tori\footnote{The precise definition is in terms of the direct integral of non-commutative torus algebras $C=\int_{\theta\in S^1} A_\theta d\theta$, with non-commutativity parameter $\theta$ varying along the base $S^1$.},
 $A_\theta \rightarrow C \rightarrow S^1$, where the non-commutativity 
 scale $\theta$ depends on the base-coordinate $x$ as
\be
\theta =  k x \,.
\ee
This proposal arose from a K-theoretical point of view by 
showing that the $H$-twisted K-theory 
of $T^3$, $K_H (T^3)$, is the same as the algebraic K-theory of the algebra associated to the field of non-commutative tori
\be
 K_H(T^3) = K (C) \,.
\ee 
 It is furthermore supported by the fact that it consistently generalizes the case of geometric fluxes and the T-duality action defined in this fashion is, thanks to Morita equivalence, of order two.
In this approach, the action of T-duality is realized in terms of taking the crossed-product 
algebra \cite{Mathai:2004qq}.

We propose that both pictures are in fact valid, and are describing
different aspects of the same T-dual compactification. More precisely,
we shall argue that the proposal (I) is the open-string version of the
T-fold proposal (II). Starting from the T-fold compactification
(\ref{TFoldData}), there is an associated open-string metric $G$ and
Theta-tensor $\Theta$ introduced and studied in
\cite{DH, CH, VS, Seiberg:1999vs}, which are related to the closed-string
 metric $g$ and $B$-field $B$ by (setting $2 \pi \alpha' =1$) 
\be \label{OpenClosed}
 G^{ij} = \left(g + B\right)^{-1}_{(ij)} \,,\qquad 
 \Theta^{ij} = \left(g +B\right)^{-1}_{[ij]} \,.  \ee 
These are the metric and spacetime
non-commutativity parameter, which the open-strings see.  For the
background in (\ref{TFoldData}) we obtain 
\be 
ds^2 = dx^2 + d\tilde{y}^2 +
d\tilde{z}^2 \,,\qquad \Theta = k x  \partial\tilde{y} \wedge \partial\tilde{z}   \,.
\ee 
This is precisely
the non-commutative torus fibration which was proposed as the T-dual
spacetime in \cite{Mathai:2004qc}. Similar backgrounds with a varying, meaning space-dependent,
non-commutativity parameter have been discussed before in
\cite{Anazawa:1999ub, Lowe:2003qy}.

How do we interpret this connection? 
The key point is to realize that the K-theory analysis depends on the open-string data (or open-string algebra). As advocated by Witten in \cite{Witten:2000cn}, the K-theory for $H$-flux backgrounds has a formulation in terms of the algebraic K-theory of a (non-)commutative algebra \cite{Bouwknegt:2000qt}, which on the other hand can be interpreted as
the open-string algebra \cite{Witten:1985cc, Witten:2000cn}. This algebra is non-commutative when $H\not=0$.
Thus in order to prove the conjectured correspondence, it remains to show that 
the algebra $C$ is precisely the algebra of open-string field theory in this background.

On more general grounds one is then led to propose the following relation: 
consider a principal $T^2$-bundle $E \rightarrow M$ 
with $H$-flux such that $H_2 \not= 0$, where $H_2 \in H^2(T^2)\otimes H^1(M)$ (``two-legged case''). 
Then the T-dual along the fibre-directions is given by a T-fold. 
The associated open-string metric and $\Theta$-tensor can be computed 
from (\ref{OpenClosed}) and the resulting space will generically be  
non-commutative, with an associated non-commutative algebra, $A$. 
The conjecture is then, that 
$A$ is precisely the algebra proposed by Mathai and Rosenberg as the T-dual, \ie, it is obtained as a crossed product algebra 
$A = C(E, H) \rtimes \mathbb{R}^2$, where
$C(E, H)$ is the $C^\ast$-algebra of the $T^2$-bundle $E$ with $H$-flux and the crossed product is taken
with respect to the $\mathbb{R}^2$-action, which is induced from the $T^2$-action on the bundle, with the K-theory of the two algebras agreeing.


\section{Probing non-geometry by generalized complex branes }

In this section the same conclusion is reached as in the last section 
by embedding the discussion into the setup of generalized complex geometry. It is shown that the T-dual of the background with $H$-flux is given by a $\beta$-transformed background. Again, this is observed in the open-string sector, and we show this by probing the T-fold geometry with generalized complex D-branes.


\subsection{Generalized complex structures, $B$-transforms and $\beta$-transforms}

Let us recall a few definitions from generalized complex (GC) geometry
\cite{Gualtieri:2003dx}.  Given an $n$-dimensional manifold $M$, a
generalized almost complex structure on $M$ is defined as an almost complex
structure on the sum of tangent and cotangent bundles $TM\oplus T^\ast
M$. For example, such a structure can be induced by an ordinary
complex structure $J$ on $M$
\begin{equation}\label{genB}\mathcal{J}_J=\begin{pmatrix}
J  & 0 \\
0 & -J^\ast \hfill 
\end{pmatrix}\,,\end{equation}
in which case it will sometimes be termed a diagonal GC structure, or by a symplectic form $\omega$ on $M$ 
\begin{equation}\label{genA}\mathcal{J}_\omega=\begin{pmatrix}
0  & -\omega^{-1} \\
\omega & 0 \hfill  
\end{pmatrix},\end{equation}
where the matrices are written in a base adapted to the direct sum.
Hybrid examples, other than these two extreme ones, are classified by
a generalized Darboux theorem \cite{Gualtieri:2003dx}, saying that any
GC space is locally the sum of a complex space and a symplectic space.
For the existence of hybrid GC structures with no underlying complex
or symplectic structure, and their relevance for $\mathcal{N}=1$
supersymmetric compactifications in string theory see \cite{Cavalcanti2005,GMT}.  For the present discussion where 
the (non-)geometry is probed by D0-branes, we shall restrict ourselves to GC
structures of the form $\mathcal{J}_J$, thus generalizing the B-model. 

Here we would like to relax the requirement that the space on which
the GC structure acts be globally of the form $TM\oplus T^\ast M$, and
we only assume that it is made of patches that look like the sum of
local tangent and cotangent spaces. The definitions are therefore to
be understood in the neighborhood of some point $p$ (which we assume
to be still well-defined), that is on $T_p M\oplus T_p^\ast M$.\\

 The sum $T_p M\oplus T_p^\ast M$ is naturally endowed with an inner
 product of signature $(n,n)$,
\be
\langle X+\xi , Y+\eta\rangle=\frac{1}{2}(\iota_X\eta+\iota_Y\xi) \,,
\ee
whose matrix in the same basis as above reads 
\begin{equation}\label{inner}
\mathcal{G}=\begin{pmatrix}
0  & 1 \\
1 & 0 \hfill 
\end{pmatrix}.\end{equation}

  The inner product is conserved by an action of the group $O(n,n)$
  whose generic element decomposes into a block-diagonal 
 part (encoding an orthogonal transformation of the tangent space and the
  induced orthogonal transformation of the cotangent space), and 
  off-diagonal blocks that can be exponentiated into $B$-transforms 
\begin{equation}\label{B-transform}\exp B =\begin{pmatrix}
1  & 0 \\
B & 1 \hfill 
\end{pmatrix},\end{equation}
\be
B : X+\xi\mapsto X+\xi+\iota_X B \,,
\ee
and $\beta$-transforms 
\begin{equation}\label{beta-transform}\exp \beta=\begin{pmatrix}
1  & \beta \\
0 & 1 \hfill 
\end{pmatrix},\end{equation}
\be
\beta : X+\xi\mapsto X+\iota_\xi \beta +\xi \,,
\ee
where $B$ and $\beta$ are antisymmetric blocks identified with a two-form
$B_{\mu\nu}$ and a bivector $\beta^{\mu\nu}$.\\

A $B$-transform acts by conjugation on generalized almost complex structures, thus  mapping the two generalized almost complex structures $\mathcal{J}_J$ and $\mathcal{J}_\omega$ to the structures 
\begin{equation}\label{JJ(B)}\mathcal{J}_J(B)=\begin{pmatrix}
J  & 0 \\
BJ+J^tB & -J^t \hfill 
\end{pmatrix}\end{equation}
and
\begin{equation}\label{Jomega(B)}\mathcal{J}_\omega(B)=\begin{pmatrix}
\omega^{-1}B  & -\omega^{-1} \\
\omega +B\omega^{-1}B & -B\omega^{-1} \hfill  
\end{pmatrix},\end{equation}
which we will encounter in section 5.


\subsection{D-branes as generalized complex submanifolds}

Let $H$ be a closed three-form. A generalized submanifold is defined
in  \cite{Gualtieri:2003dx} as a submanifold $N$ endowed with a two-form $B$ such
that $H|_N=dB$. The generalized tangent bundle $\tau_N^B$ of this
generalized submanifold is defined as the $B$-transform of the sum of
the tangent bundle $TN$  and conormal bundle (or annihilator) $\mathrm{Ann}\,TN$, namely:
\be
\label{GenTanB}
\tau_N^B=\left\{ X+\xi \in TN\oplus T^\ast M |_N, \;\xi|_N=\iota_X B
\right\}\,,
\ee
so that $\tau_N^0= TN\oplus \mathrm{Ann}\,TN$. A generalized tangent
bundle is a maximally isotropic subspace (\ie, it is isotropic with
respect to $\mathcal{G}$ and it has the maximal possible dimension for
an isotropic space in ambient signature $(n,n)$, namely $n$.)
Moreover, all the maximally isotropic subspaces are of this form, for some submanifold $N$ and two-form $B$. This
is the origin of the one-to-one correspondence between generalized
submanifolds and pure spinors, which will be used in subsection 3.4.\\

Given a GC structure $\mathcal{J}$, a generalized complex brane was
defined in \cite{Gualtieri:2003dx} to be a generalized submanifold
whose generalized tangent bundle is stable under the action of
$\mathcal{J}$. In the case of $\mathcal{J}=\mathcal{J}_J$, the
compatibility condition gives rise to the B-branes, as expected due to
the localization properties of the B-model on complex parameters
\cite{Ooguri:1996ck}.  The submanifold $N$ namely has to be a complex
submanifold, and $F$ has to be of type $(1,1)$ with respect to $J$ 
\be 
\ba
&J(TN)\subset TN \cr 
& J^\ast(\iota_X F)+\iota_{JX} F =0 \,.
\ea
\ee
In the
other extreme case of $\mathcal{J}=\mathcal{J}_\omega$, it yields all
possible types of A-branes, including the non-Lagrangian ones
\cite{Kapustin:2001ij,Chiantese:2004pe}. These are two tests of the idea that D-branes
in generalized geometries are generalized submanifolds. This idea has
passed further tests: calibrating forms and pure spinors encoding stability conditions
\cite{Marino:1999af,Kapustin:2004gv} for topological branes are
correctly exchanged by mirror symmetry
\cite{Grange:2004ah,Koerber:2005qi, BB1, BB2,LM1, LM2}, and the study of morphisms between generalized
tangent bundles \cite{Grange:2005nm} generalizes the K-theoretic
description of D-branes by taking winding numbers into account in the
resolution of vortex equations of the Yang--Mills--Higgs model
\cite{Minasian:1997mm,Witten:1998cd,Ooguri:1996ck}. Although all the
generalized tangent bundles are $n$-dimensional, a generalized
submanifold associated to a $p$-dimensional submanifold $N$ will be
sometimes referred to as a generalized D$p$-brane, and $p$ will be
called the ordinary dimension of the brane.\\

It is important for the description of D-branes in generalized
geometries to note that the projection of a subspace on the tangent
space is unchanged under a $B$-transform. A $B$-transform just
switches on an Abelian field strength with magnitude $B$ along the
brane. However, a $\beta$-transform shifts the dimension of the projection of
the brane on the tangent space (the ordinary dimension of the brane) by the rank
of $\beta$. Let us review the linear-algebraic case where the ambient
space is $V\oplus V^\ast$ for some vector space $V$. A
$\beta$-transform of a maximally isotropic subspace ${\Ann\,} F \oplus
F$, where $F$ is a subspace of $V^\ast$, reads as a graph over $F$, in
the notations of \cite{Gualtieri:2003dx}
\be
L(F,\beta)=\{X+\xi \in
V\oplus F,\, X|_F=\iota_\xi \beta\} \,.
\ee
The intersection of this space
and $V$ is just the annihilator of $F$, because it is trivially
embedded in $V\oplus F$ as 
\be
L(F,\beta)\cap V=\{X+0 \in V\oplus F,\,
X|_F=0\}=\Ann\, F\;\left(=L(F,0)\right) \,.
\ee 
The vector part of any
element of $L(F,\beta)$ therefore decomposes into an element of
$\Ann\, F$ and an element of the image of $\beta: V^\ast\rightarrow
V$, and the decomposition is unique because the graph condition
$X|_F=\iota_\xi
\beta$ implies that the intersection between $\Ann\,F$ and the image of
$\beta$ is zero-dimensional. Let $\pi_V : V\oplus V^\ast \rightarrow
V$ denote   the projection onto $V$. We have therefore argued that
\be
\pi_V L(F,\beta)=\mathrm{Im} \beta \oplus \left(L(F,\beta)\cap V\right) \,,
\ee
and therefore
\be
\mathrm{dim} \left(\pi_V L(F,\beta)\right)= \mathrm{\dim}\left(
L(F,\beta)\cap V \right) + \mathrm{rk} \beta = \mathrm{\dim} \,\Ann\, F
+ \mathrm{rk} \beta = \mathrm{\dim}\left(\pi_V L(F,0)\right) +
\mathrm{rk} \beta.
\ee
     

\subsection{T-duality maps  $B$-transforms to  $\beta$-transforms}

As we have just motivated the idea that Abelian D-branes may be
identified with GC submanifolds, and since $\beta$-transforms can
change the ordinary dimension of such submanifolds, it is natural to look for
the connection between $\beta$-transforms and monodromies on T-folds,
in the picture (II) of non-geometry. D-branes wrapped on T-folds can
come back to themselves with a different dimension after
monodromy. We are going to describe how T-dualities map $B$-transforms
to $\beta$-transforms, together with the corresponding effects on D-branes.


\subsubsection{Geometric three-torus with $H$-flux and $B$-transforms}

Consider again the flat three-torus with  uniform $H$-flux, with the same coordinates as above.
 Consider two D2-branes wrapping fibres over two points of the base, one at $x=0$ and one at generic $x$. Going from the first to the second involves a $B$-transform by the two-form 
\be 
B(x)=kx dy\wedge dz \,.
\ee
Going from $x=0$ to generic $x$ namely switches a two-form along the brane.
 The boundary conditions for open strings
 ending on a D2-brane wrapping a torus over the point $x$ (with
 embedding coordinates $X,Y,Z(\sigma,\tau)$ and the obvious notation)
 read
\be
\ba
\p_\sigma Y + kx \p_\tau Z=0,\cr
\p_\sigma Z - kx \p_\tau Y=0.
\ea
\ee


The matrix of the $B$-transform in a basis adapted to the coordinates $(y,z)$ and the dual coordinates $(\tilde{y},\tilde{z})$ reads
\begin{equation}  g= \left( \begin{array}{cccc}
1 & 0 & 0 & 0 \\
0 & 1 & 0 & 0 \\
0 & -kx & 1 & 0 \\
kx & 0 & 0 & 1\\
 \end{array} \right)
\end{equation}
    

\subsubsection{Geometric T-dual with a connection}

It is instructive to perform first the T-duality along the $y$ direction. The D2-branes wrapping the two fibres in question become D1-branes, and parametrizing the base by an angle $\theta$ with $kx=\tan\theta$, we observe that the D1-branes are rotated with respect to each other within the fibre. This reflects the fact that they now live on a torus with a connection
\be\label{}
\ba
\p_\sigma(-\sin\theta Z+\cos \theta Y)=0,\cr
\p_\tau(\cos\theta Z+\sin \theta Y)=0.
\ea
\ee
In the same basis as before, T-duality is encoded by the matrix 
\begin{equation}  g_y= \left( \begin{array}{cccc}
0 & 0 & 1 & 0 \\
0 & 1 & 0 & 0 \\
1 & 0 & 0 & 0 \\
0 & 0 & 0 & 1\\
 \end{array} \right),
\end{equation}
  and the $B$-transform is therefore replaced by one with matrix 
\begin{equation} g'= g_y^{-1} g g_y= \left( \begin{array}{cccc}
1 & kx & 0 & 0 \\
0 & 1 & 0 & 0 \\
0 & 0 & 1 & 0 \\
0 & 0 & -kx & 1\\
 \end{array} \right).
\end{equation}

Let us describe these D1-branes in terms of maximally isotropic\footnote{Isotropic is understood with respect to the inner product on the sum of the two-torus and the dual two-torus; we do not specify the embedding into $T^6$ yet; the coordinate on base only plays the role of a parameter as it is not acted on by  the T-dualities we consider.} subspaces. Start at $x=0$ with a D1-brane wrapping the $y$ circle. The corresponding pure spinor is the sum of the tangent and conormal bundles of the $y$ circle, with coordinates
\be
S^1\oplus \Ann S^1= \{y,z=0,\xi^1=0,\xi^2\} \,.
\ee
Acting on it with $g'$ yields the coordinates $(y,kx y, -kx \xi^2, \xi^2)$, which means that there are Dirichlet conditions along the one-dimensional subspace of the two-torus at $x=l$ with equation:
\be
 \tan\theta Y - Z =0 \,.
\ee
This is consistent with the fact that there is now a connection on the torus, and taking $x$ to be equal to 1 (the period of the coordinate along the base) and requiring the D1-brane to come back to itself does indeed give rise to the identification of the twisted torus 
\be
(x,y,z)\sim(x+1,y,z+ky) \,,
\ee
as it should \cite{Kachru:2002sk}.


\subsubsection{Non-geometric T-dual space and $\beta$-transforms}

Let us perform one more T-duality, along the $z$ direction, and get to
the non-geometric space. The matrix acting on the $T^2$-fibre, in going
from $x=0$ to generic $x$, in a basis adapted to the real coordinates
$(y,z,\tilde{y}, \tilde{z})$ is obtained from $g$ through conjugation by the
T-duality matrix
\[g_{yz}= \left( \begin{array}{cccc}
0 & 0 & 1 & 0 \\
0 & 0 & 0 & 1 \\
1 & 0 & 0 & 0 \\
0 & 1 & 0 & 0\\
 \end{array} \right).\]
It therefore reads
\begin{equation}\label{betatransform}
g''= g_{yz}^{-1} g g_{yz}=\left( \begin{array}{cccc}
1 & 0 & 0 & -kx \\
0 & 1 & kx & 0 \\
0 & 0 & 1 & 0 \\
0 & 0 & 0 & 1\\
 \end{array} \right),
\end{equation}
which we recognize as a $\beta$-transform by the bivector field
\be \label{betadef}
\beta(x)=kx\partial_y\wedge\partial_z\,.
\ee
 Since
$\beta$-transforms affect the vector part of maximally isotropic
subspaces, there is no way of twisting the torus to bring the D0-brane back to itself after a monodromy
around the base. Moreover, $\beta$-transforms are also associated
to open paths on the base, showing that attaching an open string to
two D0-branes sitting over different points of the base is impossible,
unless T-dualities are allowed to patch the coordinate charts together. As open strings can wind around the base before attaching themselves to the second brane, they are sensitive to the global effect of non-geometry, even if the two points on the base can be put in one single coordinate patch for the purposes of local differential geometry. It is crucial for such a global effect that the base be non-simply-connected.\\

 To sum up, T-dualities therefore relate D-branes located in different fibres. Hence they 
 are needed as changes of charts, as predicted by the proposal (II). Moreover, the transformations of the
 corresponding pure spinors  are dictated by a bivector field 
$\beta^{\mu\nu}(x)= kx \partial_y\wedge \partial_z$ depending on the coordinate along the
 base in the same way as the tensor $\theta$ of the proposal (I).\\

\subsection{Generalized D0-branes on the non-geometric T-dual}
 
As points might be disturbed by global effects in non-geometric
spaces, we would like to probe non-geometry by generalized D0-branes.
Of course, in order to be able to use techniques from generalized
geometry for describing T-duals of the three-torus with $H$-flux, we
first have to embed the three-torus into a six-torus.\\

 Let us consider a generalized B-model, and pick a complex structure
 of the form $\mathcal{J}_J$, with $J$ an ordinary complex structure
 on the six-torus. We still have a choice for the complex  structure $J$: we can
 either consider the $T^2$-fibre as an elliptic curve in this complex
 structure (which would make $B$ a tensor of type $(1,1)$ and a valid
 field strength for a D2-brane of type B wrapping the elliptic curve),
 or pick a complex structure in which $y$ and $z$ are components of
 different complex coordinates. This way $B$ would have a nonzero
 component of type $(0,2)$ and the dual torus with coordinates
 $\tilde{y}$ and $\tilde{z}$ could not support a D2-brane of the B-model.
 Let us choose the second option in order to single out the role of
 the $(0,2)$ components and their possible influence on
 non-commutativity.\\

 The way we embed the three-torus into a six-torus is therefore the
 following: the $T^2$-fibre coordinates $y$ and $z$ are real parts of
 complex coordinates $y+iy'$ and $z+iz'$, where $y'$ and $z'$ are
 coordinates along additional circles, and the base is combined with a
 third additional circle with coordinate $x'$ into an elliptic curve. In the
 sequel we shall denote the local complex coordinates we have just described by
\be\label{ComplexCoordinates}
z^1=x+ix'\,, \qquad
z^2=y+iy'\,, \qquad 
z^3=z+iz' \,.
\ee
This way $B$ is not of type $(1,1)$ and will therefore contribute
non-commutative deformations as argued in
\cite{Kapustin:2003sg}. Moreover, the $x$-dependence means that Morita
equivalence cannot be used to gauge non-commutativity away, since the
$B$-field will assume non-rational values. But for the time being, we
are interested in the effect of the $(0,2)$ and $(2,0)$ components of
the $B$-field in terms of T-duality transformations, as an
illustration of (II). The connection with non-commutativity using the
language of (I) and (III) will be made in sections 4 and 5.\\

 A few comments about the choice of GC structure are in order: we
 restrict ourselves to diagonal GC structures, thus generalizing the
 B-model. We shall see in section 4 that deformations of the
 generalized B-model are indeed sufficient to explain the connection
 between non-geometry and non-commutativity, but the reason why it is
 {\emph{a priori}} sufficient to consider a diagonal GC structure is
 that only such structures allow generalized D0-branes, which are the
 point-like objects with which one would like to probe
 non-geometry. Of course it is well-known that D-branes corresponding
 to GC structures of the form $\mathcal{J}_\omega$ do not include
 D0-branes, moreover the generalized Darboux theorem implies that
 hybrid GC structures locally have some A-type boundary conditions
 that forbid D0-branes.\\

Let us consider generalized D0-branes for the GC structure we have
just described, and the way they transform under monodromy. They are
not affected by $B$-transforms because the graph condition of
definition (\ref{GenTanB}) is empty. On the other hand, their ordinary
dimension is raised upon a $\beta$-transform by an amount equal to the
rank of $\beta$, as was explained above.\\

 In order to work out the effect of the monodromy on D0-branes, we are
 going to use the description of generalized tangent bundles by pure
 spinors.  The mapping between isotropic spaces and spinors is made
 manifest by the action of sections of $TM\oplus T^\ast M$ on
 $\Lambda^\bullet M$, which carries a representation of
 ${\mathrm{Clifford}}(n,n)$:
 \begin{equation} (X+\xi).\phi=\iota_X \phi+ \xi\wedge\phi,\end{equation}
 \begin{equation} (X+\xi).((X+\xi).\phi)= \langle X+\xi, X+\xi\rangle
\phi.\end{equation}
 Given a spinor, one can associate to it its null
 space in $V\oplus V^\ast$. Maximally isotropic subspaces are
 therefore in one-to-one correspondence with pure spinors.\\

 In the case of a generalized D0-brane, the pure spinor to be
 considered is the holomorphic three-form
\be
\Omega:=dz^1\wedge dz^2\wedge dz^3 \,,
\ee
in a local patch where $z^1,z^2,z^3$ are complex coordinates associated to the complex structure we have described on the six-torus. The annihilator is
locally of the form  
\be
TM^{(0,1)}\oplus T^\ast M^{(1,0)}
  =\Vect\left( \frac{\p}{\p\bar{z}^1},\frac{\p}{\p\bar{z}^3},
       \frac{\p}{\p \bar{z}^3}\right)\oplus \Vect\left(dz^1, dz^2, dz^3 \right)\,.
\ee
 
 Let us write the components of $\beta$ in a way adapted to the local
 complex coordinates, so that $\beta^{\mu\nu}$ is the $(-2,0)$ part of
 $\beta$, and $\beta^{\bar{\mu}\bar{\nu}}$ and $\beta^{\mu\bar{\nu}}$
 do not appear in the $\beta$-transform because they act on components
 of the annihilator that are zero (and stay so, because the one-form
 part is not transformed by $\beta$). The transformation rules are
 therefore
\be\label{betamono1}
\xi^{\bar{\mu}} \bar{\partial}_\mu +\xi_\mu dz^\mu\longrightarrow (\xi^{\bar{\mu}})' \bar{\partial}_\mu+ ({\xi}^\mu)'\partial_\mu+ \xi_\mu dz^\mu,\ee
where
\be\label{betamono2}
\ba
(\xi^{\bar{\mu}})' & =\beta^{\bar{\mu}\nu}\xi_\nu+ \xi^{\bar{\mu}}\,,\cr
(\xi^\mu)'         & =\beta^{\mu\nu}\xi_\nu \,.
\ea
\ee
 The vector space spanned by the
vectors $(\xi^{\bar{\mu}})'\p_{\bar{\mu}}$ is still the whole subspace
$TM^{(0,1)}$, whereas the projection of the annihilator of $\Omega$ on
$TM^{(1,0)}$ is made two-dimensional by the monodromy, since $\mathrm{rk}\,\beta^{\mu\nu}=\mathrm{rk} (dz^2\wedge dz^3)=2$.\\

 As expected, the dimension of the projection on the tangent space is
 shifted by the rank of the $(0,-2)$ part of the bivector field
 $\beta$. This establishes that there is no zero-dimensional global
 section of the vector part.  This phenomenon was observed in the
 context of non-commutative deformations in \cite{Kapustin:2003sg},
 where it was called the uncertainty principle for topological
 D-branes (commutators of equations of complex submanifolds cannot
 vanish, and this prevents D-branes from wrapping maximal-codimension
 submanifolds). In the present case, the change of type\footnote{A
 pure spinor can be written in a unique way as
 $\theta^1\wedge\dots\wedge\theta^n \wedge e^F$, where
 $\theta^1,\dots,\theta^n$ are complex one-forms and $F$ is a complex
 two-form; the integer $n$ is called the {\emph{type}} of the pure
 spinor.} of a pure spinor under monodromy is equivalent to the lack
 of a global splitting between momenta and winding numbers. The space
 obtained by T-duality from the generalized complex $T^6$ with
 $H$-flux can therefore not be globally of the form $L\oplus L^\ast$,
 with $L$ a maximally isotropic subspace. This is the absence of
 global polarization that appeared in the real case for T-folds, and
 it is encoded by the $(0,-2)$ part of the bivector field $\beta$.\\


\section{Non-commutativity from Lagrangian deformations in the BV procedure}

 In the previous section our consideration of the T-dual of a complex
 six-torus with $H$-flux has shown the necessity of T-folds for the
 description of the global topology, when no global polarization
 exists, which reproduces the criterion of \cite{Hull:2004in} for the
 objects (II). What about proposal (I) and non-commutativity along the
 dual $T^2$-fibres? The connection comes from the construction of
 topological string theory on GC spaces in \cite{Pestun:2006rj}, where
 it was explained that a tensor $\beta$ of type $(0,-2)$ can deform
 the B-model with generalized complex target space, inducing
 star-products on the fibre. In order to make contact with
 \cite{Mathai:2004qq} we are going to show how the $\beta$-transform
 induces this very deformation of the generalized B-model on the
 T-dual of the GC six-torus. For a categorical viewpoint on the
 Fourier--Mukai equivalence between deformations of complex tori
 (either in a non-commutative direction parametrized by a holomorphic
 Poisson structure or in a $B$-field direction), see
 \cite{Ben-Bassat:2005ne}.


\subsection{$\beta$-transforms and the generalized B-model}

The Batalin--Vilkovisky (BV) formalism requires a nilpotent operator $Q$ and an odd differential operator of second order $\Delta$. They act on the graded space of fields and induce an odd symplectic structure, for which the master action $S$ is a Hamiltonian function. The odd Laplacian $\Delta$ induces an antibracket via the formula
\be
(F,G)=(-1)^{|F|}\Delta(F\wedge G) -\Delta F\wedge G- (-1)^{|F|}F\wedge \Delta G \,.
\ee
The condition $Q^2=0$ then induces the master equation
\be
(S,S)=0 \,.
\ee

 From now on, as is required by the BV procedure, we shall give
 fermionic statistics to the vector and form coordinates, or in other
 words reverse the parity on the fibres of the tangent and cotangent
 bundles. When computing the action of a sigma model, one has to pull
 back vector and form fields on the world-sheet $\Sigma$, which
 induces a change of statistics on the tangent bundle of the
 world-sheet, which is now denoted $\hat{\Sigma}$.  As far as the
 B-model is concerned, the graded space of fields is (in a local
 coordinate patch) the space of observables of the B-model. The
 operators $Q$ and $\Delta$ are the anti-holomorphic and holomorphic
 differentials,
\be
\ba 
 Q &= \bar{\partial}=d\xi^{\bar{\mu}}\frac{\partial}{\partial
 \bar{\xi}^{\bar{\mu}}} \cr
 \Delta &=\partial=dz^\mu\frac{\partial}{\partial \xi^\mu}=
 \frac{\partial}{\partial {\xi}_\mu}\frac{\partial}{\partial
 {\xi}^\mu} \,,
\ea
\ee
 where in re-expressing $\partial$ as a second-order
 differential operator, use has been made of the observation that
 one-forms may act on the de Rham complex as derivatives with respect to
 vector coordinates \cite{Witten:1990wb}. This way the coordinates
 $\xi_\mu$ and $z^\mu$ are canonically conjugate to each other, and
 one has to add antifields to be paired with $\bar{z}^{\mu}$ and
 $\xi^{\bar{\mu}}$ (because $\Delta$ is degenerate on the subspace
 they span), called $\bar{z}^\ast_{\mu}$ and
 $\xi^\ast_{\bar{\mu}}$. As shown in \cite{Pestun:2006rj}, the master
 action for the generalized B-model then reads
\be
S=\int_{\hat{\Sigma}}\left(\xi_\mu
dz^\mu+\xi^{\bar{\mu}}z^\ast_{\bar{\mu}}\right) \,.
\ee

 The allowed deformations of the generalized B-model involve holomorphic bivector fields.
 A Lagrangian submanifold $L$ of the space of fields has indeed to be
chosen to compute the gauge-fixed partition-function
\be
Z_L:=\int_L DX e^{-S[X]} \,,
\ee
and the invariance of this path integral
under change of the gauge-fixing condition is equivalent to its
invariance under the Lagrangian deformations of $L$. A variation in
the gauge-fixing condition amounts to a Lagrangian deformation of the
Lagrangian submanifold $L$, namely one where the momenta are derived from a density
\be
\delta p^i=\frac{\p \Xi}{\p x^i} \,.
\ee
Starting with a Lagrangian
submanifold with equation given by the vanishing of all momenta
$p^i=0$, invariance of $Z_L$ under Lagrangian deformations is
expressed by the following chain of equalities
\be
\delta_L Z =\int_{L}\delta p^i \frac{\delta }{\delta p^i}{e^{-S(x_i)}}=\int_L
\frac{\p \Xi}{\p x^i} \frac{\delta }{\delta p^i}{e^{-S(x_i)}}=-\int_L
\Xi \frac{\p}{\p p_i}\frac{\p}{\p x^i}\left({e^{-S(x_i)}}\right)=0 \,,
\ee
which implies 
the quantum master equation
\be
\Delta(e^{-S[X]})=0 \,.
\ee
Expanding in powers of a deformation of the master action gives rise to the Maurer--Cartan equation.
 In the case of the generalized B-model, splitting into
 tensors of different types shows \cite{Pestun:2006rj} that the deformation
 of the generalized B-model by a holomorphic bivector field is allowed
 (moreover, the sum of tangent and cotangent spaces is one of the geometries recently addressed by Ikeda in the deformation theory of BV structures \cite{Ikeda:2006wd}).\\

In the present context, the lack of global polarization induces
deformations of the BV structure when going from one patch of
coordinates to antother.  It is instructive to see how derivatives are
affected by the monodromies described in (\ref{betamono1}).  Let us
work out the deformation of the antibracket adapted to the isotropic
subspace $TM^{(0,1)}\oplus T^\ast M^{(1,0)}$ we started with in the
previous section 
\be
(F,G)=F\frac{\p}{\p \xi^\mu}\frac{\p}{\p\xi_\mu}G
- F\frac{\p}{\p\xi_\mu}\frac{\p}{\p \xi^\mu} G \,.
\ee
The change of
coordinates induced by a $\beta$-transform
\be\label{}
\ba
\xi'^\mu=\beta^{\mu\nu}\xi_\nu +\xi^\mu\,,\cr
\xi'^\mu=\beta^{\mu\nu}\xi_\nu +\xi^\mu\,,
\ea
\ee
induces the following changes in derivatives on the cotangent space
\be
\ba
\frac{\p}{\p \xi^\mu}   
     &=     \frac{\p}{\p {\xi'}_\nu}\frac{\p{\xi'}_\nu}{\p \xi^\mu}+
            \frac{\p}{\p {\xi'}^\nu}\frac{\p{\xi'}^\nu}{\p \xi^\mu}
     =      \frac{\p}{\p {\xi'}^\mu}\,, \cr
\frac {\p}{\p \xi_\mu}
    &= \frac{\p}{\p {\xi'}_\nu}\frac{\p{\xi'}_\nu}{\p \xi_\mu}
          + \frac{\p}{\p {\xi'}^\nu} \frac{\p{\xi'}^\nu}{\p \xi_\mu}
     =\frac{\p}{\p {\xi'}_\mu}+\beta^{\mu\nu} \frac{\p}{\p{\xi'}^\nu}\,,
\ea
\ee
and the antibracket now includes pairs of derivatives with respect to the vector
coordinates, so that the monodromy shifts the antibracket by a Poisson bracket:
\be
(F,G)'=(F,G)+2 F \frac{\p}{\p {\xi'}^\mu}\beta^{\mu\nu}\frac{\p}{\p {\xi'}^\nu} G \,.
\ee
 We have therefore shown that the data of the  BV structure do change from patch to patch in the
 non-geometric T-dual of the GC six-torus. We are going to work
 directly on the master action, since the $\beta$-transform is of the
 Lagrangian type, so that the T-duality that has been seen to bind
 together the coordinate patches, is also deforming the master action.


\subsection{From the generalized B-model to the Poisson sigma model}

  We have seen in the previous section that in a special case an
  obstruction to the existence of a global generalized complex form $L
  \oplus L^\ast$ for the T-duals is encoded by a holomorphic bivector
  field $\beta^{\mu\nu}$. The link with \cite{Mathai:2004qc} is
  provided by the choice of an isotropic submanifold involved in the
  BV gauge-fixing procedure. A global such choice is impossible as
  soon as the $\beta$-transform is non-trivial, and this leads to a
  deformation of the generalized B-model by the holomorphic bivector
  field $\beta$.  The resulting model is precisely the Poisson sigma
  model that appears as the $\beta$-deformation of the topological
  $\mathcal{J}$-model constructed by Pestun
  \cite{Pestun:2006rj}. Star-products emerge from the Poisson
  structure by deformation quantization
  \cite{Schaller:1994es,Kontsevich:1997vb}. Of course this is no
  accident. Relevance of the Kontsevich formula in non-commutative
  gauge theory along D-branes appeared for example in
  \cite{VS,Cornalba:1999ah,Jurco:2000fb}.
 
 Consider the master action that is obtained from the BV procedure for
 the B-model with a generalized complex manifold as a target space
 \cite{Pestun:2006rj}, \ie, a target space endowed with a GC structure of the
 diagonal form $\mathcal{J}_J$. We therefore start with the master action on the
 patch with complex coordinates $(z^\mu,\bar{z}^\mu)$
\be
S =\int_{\hat{\Sigma}}\left(\xi_\mu dz^\mu+\xi^{\bar{\mu}}z^\ast_{\bar{\mu}}\right).
\ee
 As for the $\xi^{\bar{\mu}}$, they span the bundle $TM^{(1,0)}$ that is not
modified by the monodromy, and the $z^\ast_{\bar{\mu}}$ are conjugate
to the antiholomorphic base coordinates that are untouched by the
monodromy. The second term is decoupled in the initial patch, and will
stay so under monodromy.

 But the first term, as it is endowed with  holomorphic indices, is affected by the
 monodromy. This is due to the fact that the circle base cannot be covered by a single patch. Let
 us choose a construction of the spinor bundle where the differential
 forms act by differentiation with respect to the dual coordinates. This
 corresponds to choosing the pure spinor $\Omega$ as the vacuum, and
 vector fields as creation operators, as explained by Witten in
 \cite{Witten:1990wb}. In a local patch we therefore identify $dz^\mu$ with 
$\p/\p\xi_\mu$, so that the relevant term in the master action transforms as follows
\be
\left( \xi_\mu\frac{\p}{\p\xi_\mu}\right)'
   = \xi_\mu \frac{\p}{\p\xi_\mu}+\xi_\mu\beta^{\mu\nu}\frac{\p}{\p\xi^\nu},
\ee 
and the result, in a representation where differential forms act by
multiplication, as
\begin{equation}\label{patches} 
{S'} =\int_{\hat{\Sigma}} \left( \xi_\mu dz^\mu+\beta^{\mu\nu}\xi_\mu
\xi_\nu\right)\,,
\end{equation} 
which of the form $S+\delta_\beta S$, with $\delta_\beta S$ induced by the holomorphic bivector $\beta$. Of
course we can rewrite the expression for $S'$ in terms of coordinates,
vectors and forms with conventional statistics, both on the world-sheet
$\Sigma$ and the target space, by taking multiplications to be wedge
products.

Since the non-zero components of the bivector field $\beta$ are along the directions $y$ and $z$, and they depend only of the coordinate $x$, the parameter $\beta$ is a Poisson bivector field
\be
\beta^{\mu\nu} \partial_{\nu} \beta^{\rho\sigma} +\beta^{\rho\nu} \partial_{\nu} \beta^{\rho\mu} +\beta^{\sigma\nu} \partial_{\nu} \beta^{\mu\rho} =0\,. 
\ee
The resulting model with action $S'$ is the Poisson sigma model
studied by Cattaneo and Felder in \cite{Cattaneo:1999fm}. We are therefore left with T-dual  fibres forming a field of non-commutative tori over a circle. A subspace of the T-dual $T^6$ is therefore non-commutative, with the non-commutativity scale predicted by (I). \\

For a given topology of the world-sheet, we
 find a deformation of the product of observables into a
star-product $\ast_\beta$ as in \cite{Cattaneo:1999fm}:
\be
f\ast_{\beta(a_1)} g (a)=\int_{X_{\infty}=a} DX f(X(0))g(X(1)) e^{i(S+\delta_\beta S)[X]} \,,
\ee
where $0,1,\infty$ are the coordinates of the points of insertion of observables on the boundary of the world-sheet, and $a_1$ is the component of the coordinates of the point $a$ along the direction $x$. The continuous dependence on $a_1$ comes from the definition (\ref{betadef}) for the bivector $\beta$. The fact that non-commutativity shows up in the boundary correlators is the sign that the open-string sector is crucial for the equivalence between (I) and (II).


\section{$SU(3) \times SU(3)$ structure and non-commutativity}

It is argued in \cite{GLW} that the mirror of a Calabi--Yau
compactification with magnetic $H$-flux \cite{Gurrieri:2002wz}
possesses an $SU(3)\times SU(3)$ structure. $SU(3) \times SU(3)$
structure compactifications are described in terms of pure spinors,
made from bilinears of $SU(3)$-invariant spinors of
$\mathrm{Cliff}(6)$. In case those invariant spinors are not parallel
and the type of the associated pure spinors is not globally defined, 
the resulting compactification is conjectured to be
non-geometric. In case the two pure spinors still have a
globally constant type, the compactification has global $SU(2)$
structure and is still geometric, an example of which is $T^2 \times
K3$.  The proposal in \cite{GLW} should in particular be consistent
with the non-commutative T-dual conjecture, when both setups are
applicable. \\

In this section we show that precisely in the case when  the two $SU(3)$-invariant spinors are not parallel, there is a non-trivial Poisson bivector, which yields a non-commutative 
deformation of the open-string background. If the pure spinors have a uniform expression, the Poisson bivector 
is constant and thus one can dispose of the non-commutativity by Morita equivalence. This of course is 
not possible in the case when the Poisson bivector is dependent on the remaining coordinates.


\subsection{$SU(3) \times SU(3)$ structure manifolds}

Consider Type II compactifications on six-manifolds with $SU(3)\times SU(3)$ structure \cite{Hitchin:2004ut,Gualtieri:2003dx, Fidanza:2003zi,Grana:2005sn,Grana:2005ny,Benmachiche:2006df}
(for a detailed 
introdution and references see \cite{Grana:2005jc}). 
As such they are characterized by a pair of no-where vanishing $SU(3)$-invariant spinors $\eta^{1,2}$, 
which arise in the decomposition of the two $SO(9,1)$ spinors $\epsilon^{1,2}$ of Type II under 
$SO(3,1) \times SO(6)$. 
If $\eta^1 =\eta^2$ the structure group is reduced to $SU(3)$, which in particular incorporates the case of standard Calabi--Yau compactifications. 
If on the other hand the spinors are not parallel to each other throughout the manifold, one speaks of an $SU(2)$ structure. The latter is characterized by a non-vanishing vector field. Defining
\be\label{EtaOneTwo}
\eta_+^2 = c \eta_+^1 + (v + i w) \eta_-^1 \,,\qquad c\in \mathbb{C}\,,
\ee
the vector field in question is 
\be\label{OneForm}
\nu_m:= {\eta_{+}^1}^\dagger \gamma_m \eta_-^2 = v_m - i w_m \,.
\ee
The spinors $\eta^{1,2}$ can also be combined to construct a pair of $SU(3,3)$ bi-spinors
\be
\Phi_{\pm} = \eta_+^{1} \otimes {\eta_{\pm}^2}^\dagger \,,
\ee
which are pure (\ie, they are annihilated by half of the $\Gamma$-matrices). Moreover,  
via the standard Clifford map, they  are in one-to-one correspondence with (formal sums of) differential forms.
On an $SU(3)$ structure manifold the pure spinors $\Phi_\pm$ correspond to the $(3,0)$ form $\Omega$ and the (exponential of the) $(1,1)$ form $J$.
Generically however, there are two independent two- and three-forms
\be
\ba
J_{\pm}      &= j \pm v \wedge w \cr
\Omega_{\pm} &= \omega \wedge (v \pm i w) \,.
\ea
\ee
Here $j$ and $\omega$ parametrize the local $SU(2)$ structure in the transverse directions to $v$ and $w$.
For the present purposes it is instructive to note that in the case of non-geometric spaces the notion of transversality holds only locally, since it can be spoiled by a $\beta$-transform. 

The associated two pure spinors are
\be
\Phi_+ =  {1\over 8} (\bar{c}  e^{- i j} - i \omega) \wedge e^{- i v \wedge w} \,,\qquad
\Phi_- = - {1\over 8} (e^{- i j} + i c \omega) \wedge (v + i w) \,.
\ee
Raising an index on the two-forms $J_\pm$ we obtain two complex structures, $I_\pm$, which on the other hand define a generalized complex structure $\mathcal{I}$ as will be discussed in the next section.
Thus, one important point to notice is that $SU(3) \times SU(3)$ structure implies generically that 
there are two independent complex structures, which arise from 
the two $SU(3)$-invariant spinors $\eta^{1,2}$.


\subsection{Non-commutative deformations}

To begin with, let us review some known facts about non-commutativity:
In \cite{Kapustin:2003sg} Kapustin presents a criterion when a compactification allows for non-commutative deformations. Consider first the case of closed $B$. Then for a Calabi--Yau manifold with metric $G$ and $B$-field and two (not necessarily equal) complex structures $I_{\pm}$ compatible with the Levi--Civita connection, one may define the generalized complex structure
\be
\ba
\label{Idef}
 \mathcal{I} &= \left(
   \begin{array}{cc}
     \half(I_+ + I_-) + \delta P\, B & - \delta P \cr
     \delta J + B \delta P \, B 
       + \half B(I_+ + I_-) +  \half (I_+ + I_-)^t B \quad  
        &  - \half  (I_+ + I_-)^t - B \delta P 
   \end{array}
               \right) \,,  
\ea
\ee
where the bivector part is defined as
\be
\delta P         = \half( J_+^{-1} - J_-^{-1}) \,,
\ee
with $J_{\pm} = G I_{\pm}$. Furthermore $\delta J = 1/2 (J_+ - J_-)$.
The complex structure is
\be
I =  \half(I_+ + I_-) + \delta P\, B \,,
\ee
whereas the Poisson bivector, which parametrizes the non-commutative deformations, is given by
\be
\theta = -\half I \delta P \,.
\ee
In particular, the non-commutativity is non-trivial only if the two complex structures 
are unequal $I_+ \not= I_-$, since otherwise $\delta P =0$. 
In \cite{Kapustin:2004gv} this was generalized to the case of $H=dB \not= 0$. The only 
difference from the above equations is that the complex structures $I_\pm$ now have to be covariantly constant 
with respect to the connection with  torsion $H_{ij}\,^k$.
Note however that as in the T-fold case, non-commutativity arises only at the level of D-branes, 
\ie, the closed-string background remains commutative, albeit not necessarily geometric, as the left- 
and right-moving modes on the world-sheet differ.  In particular, a
generic $SU(3)\times SU(3)$ structure compactification yields a pair
of distinct complex structures, and thus two distinct realizations of
the $\mathcal{N}=2$ super-conformal algebra for left- and
right-movers, respectively. In this sense, the world-sheet theory is
very much alike the situation for asymmetric orbifolds.\\

In particular, we can then determine $\delta P$ in the case of $SU(3)\times SU(3)$ structure compactifications
\be
\delta P = - \half ((GI_+)^{-1} - (GI_-)^{-1}) \,,
\ee
which has again non-vanishing Poisson bivector $\theta = -1/2 I \delta
P$ if the two complex structures differ.

In case of the two spinors being never parallel, which corresponds to $v+iw\not=0$, \ie, we have an $SU(2)$ structure at least locally, the corresponding $\theta$ is non-zero. So this is indeed the case, when there are non-trivial non-commutative deformations. In fact we can write the Poisson bivector entirely in terms of the one-form $\nu$ that characterizes the $SU(2)$ structure
\be
\delta P= w \wedge v = {1\over 2 i} \nu \wedge \bar{\nu}\,,
\ee
which means that the non-commutativity is governed only by the vectors in (\ref{OneForm}). 

If the $SU(2)$ structure is global, such as for $K3$-compactifications, the resulting non-commutative deformations are constant and thus of minor interest to the present discussion. The interesting cases
arise, when the above description is only local. Then the Poisson bivector is not constant and one cannot get rid of it by Morita equivalence.  
Thus, the lack of global definition for the type of the pure spinors (because of the change of dimension between two different base-points that was illustrated above) makes it unlikely that the 
linear independence between two $SU(3)$ spinors can be described by a globally-defined vector field. 
We may obtain a Poisson bivector $\delta P$, but it need not be of the form $v\wedge w$. We shall present an explicit example in the next section.


\subsection{Torus with $H$-flux and mirror symmetry}

An illustrative example is $T^6$ in the complex coordinates (\ref{ComplexCoordinates})
with $H$-flux $H\in H^3(T^6, \mathbb{Z})$, which allows to 
use the language of $SU(3)\times SU(3)$ structures. 
We consider a triple T-duality along $x', y, z$, which in this context corresponds to considering the mirror. The three-form is $\Omega = dz^1 \wedge dz^2 \wedge dz^3$ and the $(1,1)$-form is $J= \sum_i dz^i \wedge d\bar{z}^i$.
In particular, the $T^3$ that the T-duality acts upon is a special Lagrangian cycle. 
Note that this is different from the T-duality transformations encountered in the previous sections for $T^3$ with $H$-flux, however, as in that case, we consider only two-legged $H$-flux, \ie, 
T-duality acts in only two directions supporting the $H$-flux. 

The setup for the square torus is simple enough, and the generalized complex structure $\mathcal{I}$ is diagonal. However switching on the $H$-flux yields
\be
 \mathcal{I}(B) = \left(
   \begin{array}{cc}
     I  & 0 \cr
     BI+I^tB & -I^t 
   \end{array}
               \right) \,,  
\ee
The T-dual complex structure was determined in \cite{Kapustin:2003sg} to be 
\be
 \mathcal{I}'(B) = \left(
   \begin{array}{cc}
     - I^t  & B I + I^t B \cr
     0 & I 
   \end{array}
               \right) \,, 
\ee
so that the Poisson bivector is read off to be $\delta P=  B I + I^t B$, which is not necessarily vanishing, as expected. Moreover, if we insist that the $\beta$-transformed  D0-brane be a generalized D0-brane with respect to   $\mathcal{I}'$ after monodromy, we obtain the constraint $\beta=\beta^{(0,2)}$, in terms of the decomposition with respect to $I$. The deformation parameter is once more seen to be a (0,-2) tensor.\\

 Let us connect this non block-diagonal GC structure to the language
 of maximally isotropic subspaces we used to probe the non-geometry
 by D0-branes. Consider again a graph of some bivector field $\beta$
 over some subspace $F$ of the cotangent space
\be 
L(F,\beta)=\{X+\xi\in V\oplus F,\,  X|_F=\iota_\xi\beta\} \,,
\ee
 and require
 stability of this graph under the action of $\mathcal{I}'$. We are led
 to the following equation, that must hold for every element $\xi$ in
 $F$
\be
-(I^t)^\mu_\nu(\beta^{\nu\rho}\xi_\rho)+(\delta P)^{\mu\nu}\xi_\nu=\beta^{\mu\nu} I_\nu^\rho \xi_\rho \,,
\ee
Eliminating
 the coordinates $\xi_\rho$ we observe that the $(0,-2)$ part of
 $\beta$ must equal the Poisson bivector field
\be
\beta I+I^t \beta=  \delta P \,.
\ee
We therefore see that the non-diagonal block of the GC
 structure in the T-dual picture is precisely the parameter of the 
 $\beta$-transform that is undergone by any D0-brane. Whenever $\delta
 P$ is non-zero, the dimension of the projection of a D-brane onto the
 tangent space is non-zero. Therefore $\delta P$ induces non-geometry
 in the sense that point-like D-branes cannot be put on a GC space
 with non-zero bivector block.\\

 As stated in theorem (5.4) of \cite{Kapustin:2000aa} for mirrors of
 complex tori, the two generalized complex structures
 ${\mathcal{J}}_J (B)$ and ${\mathcal{J}}_\omega (B)$ shown in formulae (\ref{JJ(B)}) and (\ref{Jomega(B)}) are exchanged by mirror symmetry. Indeed, if $g_{x'yz}$ is the element of $O(6,6)$ encoding T-duality in the $x',y,z$ directions, the mirror exchange
\be
\left({\mathcal I}(B)\right)'=g_{x'yz}\mathcal{J}(B)g_{x'yz}^{-1} \,,
\ee
holds.\\ 

 As we already argued, the image of ${\mathcal{J}}_J$ by T-duality is not block-diagonal anymore, and we may read off the Poisson bivector $\delta P$ as
\be
\delta P= \frac{1}{2}kx\left(\partial_{y'} \wedge \partial_{z}+ \partial_{y} \wedge \partial_{z'} \right)\,,
\ee
which is not decomposable as the tensor product of two vectors. Hence this is a case of generalized type of T-fold. In this example of a non-geometric T-dual, the non-commutativity scale 
 depends on the base coordinate $x$ in a way that prevents to gauge it away by Morita equivalence, and furthermore it does not come from a globally defined vector field that encodes the linear independence between two $SU(3)$-invariant spinors. The lack of a global polarization in T-folds (II) can therefore be traced in the formalism of (III) as the non-decomposability of the Poisson bivector field $\delta P$.

 So far our consideration of the T-dual of a complex torus with
 $H$-flux has shown the necessity of T-folds for the description of
 the global topology, when no global choice of type for a pure spinor exists,
 which reproduces the criterion of \cite{Hull:2004in}. What about
 non-commutativity along the dual $T^2$-fibres? The connection comes
 from the construction of topological string theory on GC spaces and sigma models with bistructures in
 \cite{Zucchini:2004ta,Zucchini:2005rh,Pestun:2006rj,Howe:2005je}. It was explained that a tensor $\beta$  of type $(0,-2)$ can deform the B-model with generalized complex target space, yielding star-products induced by the Poisson bivector
 $\beta$. In order to make contact with \cite{Mathai:2004qq} it would be interesting
 to see how the $\beta$-transform induces this very same deformation
 of the generalized B-model.

The results in this section should also be derivable from the action of T-duality on spinors as advocated by Hassan \cite{Hassan:1994mq, Hassan:1999bv, Hassan:1999mm}. In the geometric case, in particular for the the Lunin--Maldacena background \cite{Lunin:2005jy}, the analysis was performed in \cite{Minasian:2006hv} and it is easily observed from their results that $\delta P$ in this case is of the form $v\wedge w$, and non-commutativity does not occur between coordinates but as relative phases between ordered products of fields, corresponding to global $U(1)$ symmetries.

 
\section{Conclusion}

T-duality in the presence of NSNS fluxes provides the first stepping stone to understanding 
generalized versions of mirror symmetry \`a la Strominger--Yau--Zaslow (SYZ) \cite{Strominger:1996it}. 
The present paper discusses this issue, thereby merging various existing proposals for the T-dual.  
If the NSNS $H$-flux is supported only on one T-dualized direction the dual is again geometric and 
consensus has been reached on the T-dual geometry throughout the literature. Controversy starts when two  T-dualized directions are spanned by the $H$-flux. Our present investigations concern the case of two directions, 
applied to tori and torus fibrations. 
We have shown that the proposal (I) of Mathai and Rosenberg, claiming the dual to be a  field of non-commutative tori, can be viewed as the open-string version of Hull's T-fold proposal (II). Secondly, we have shown that 
generalized geometries provide an alternative setup for studying the T-dual or mirror, and can be reconciled 
with the non-commutativity proposal by explicit construction of a Poisson bivector, which depends crucially on 
the background $H$-flux. As argued in \cite{Kapustin:2003sg,Grange:2004ra}, this bivector parametrizes non-commutative deformations of the 
open strings.
We found that in the case of two-legged $H$-flux, the bivector field is in fact not uniform, but varies along 
the base of the torus-fibration.\\

On more general grounds it would be interesting to understand the precise conditions for the 
dual space to be non-geometric. 
The key ingredient for the deformation by the bivector field is
of course the multiple-connectedness of the base of the fibration. The
present discussion could be extended to complicated fibrations over a
multiply-connected base with $H$-flux. On the other hand, T-dualizing
along a two-torus carrying a non-zero $B$-field and fibered over a
contractible space, as in the sequence of $\beta$-deformations in
\cite{Lunin:2005jy}, can lead to a geometric T-dual (orbifolds with torsion in
that instance). This is consistent with the fact that any loop on the
base can be shrunk and included in a local coordinate patch (thus
removing the $\beta$-transformed term in the formula (\ref{patches}) in
the simply-connected cases), and also with the fact that
non-commutativity of the dual two-torus in the proposal of
\cite{Mathai:2004qq} is measured by classes in the first integral cohomology
group of the base (thus setting the star-product to the ordinary
product in the simply-connected cases).\\

Clearly it would be very interesting to extend the present discussions to more general setups, 
in particular, a generalized SYZ construction would be the most natural next problem to be addressed. 
 The argument of SYZ for the existence of
 $T^3$-fibrations of Calabi--Yau manifolds rested on the
 fact that a D0-brane had only its position as a modulus. In the case
 of T-folds, the modulus $\beta$ is a modulus of Lagrangian
 deformations and prevents D0-branes from existing, just as
 non-commutativity does.


\subsection*{Acknowledgments}

We thank Mariana Gra$\tilde{{\rm n}}$a, Simeon Hellerman, Jan Louis, Varghese Mathai, Ruben Minasian, 
Bastiaan Spanjaard and Dan Waldram for discussions and comments on various occasions. P.G. thanks the Institute for Advanced Study for hospitality, as well as the organizers of the 38th International Symposium Ahrenshoop for a stimulating conference, and is funded by the German-Israeli Foundation for Scientific Research and Development. S.S.N. thanks the Erwin Schr\"odinger Institute, Wien, for hospitality and is funded by a Caltech John A. McCone Postdoctoral Fellowship in Theoretical Physics. 
This work was supported in part by the DFG and the European RTN Program MRTN-CT-2004-503369.



\newpage


\bibliographystyle{JHEP} \renewcommand{\refname}{References}

\begin{thebibliography}{10}

\bibitem{Alvarez:1993qi}
E.~\'Alvarez, L.~\'Alvarez-Gaum\'e, J.~L.~F. Barb\'on, and Y.~Lozano, {\it Some
  global aspects of duality in string theory},  {\em Nucl. Phys.} {\bf B415}
  (1994) 71--100, [\href{http://xxx.lanl.gov/abs/hep-th/9309039}{{\tt
  hep-th/9309039}}].

\bibitem{Gurrieri:2002wz}
S.~Gurrieri, J.~Louis, A.~Micu, and D.~Waldram, {\it Mirror symmetry in
  generalized {Calabi--Yau} compactifications},  {\em Nucl. Phys.} {\bf B654}
  (2003) 61--113, [\href{http://xxx.lanl.gov/abs/hep-th/0211102}{{\tt
  hep-th/0211102}}].

\bibitem{Kachru:2002sk}
S.~Kachru, M.~B. Schulz, P.~K. Tripathy, and S.~P. Trivedi, {\it New
  supersymmetric string compactifications},  {\em JHEP} {\bf 03} (2003) 061,
  [\href{http://xxx.lanl.gov/abs/hep-th/0211182}{{\tt hep-th/0211182}}].

\bibitem{Hellerman:2002ax}
S.~Hellerman, J.~McGreevy, and B.~Williams, {\it Geometric constructions of
  nongeometric string theories},  {\em JHEP} {\bf 01} (2004) 024,
  [\href{http://xxx.lanl.gov/abs/hep-th/0208174}{{\tt hep-th/0208174}}].

\bibitem{Fidanza:2003zi}
S.~Fidanza, R.~Minasian, and A.~Tomasiello, {\it Mirror symmetric
  {$SU(3)$}-structure manifolds with {NS} fluxes},  {\em Commun. Math. Phys.}
  {\bf 254} (2005) 401--423,
  [\href{http://xxx.lanl.gov/abs/hep-th/0311122}{{\tt hep-th/0311122}}].

\bibitem{Bouwknegt:2003vb}
P.~Bouwknegt, J.~Evslin, and V.~Mathai, {\it T-duality: Topology change from
  {H}-flux},  {\em Commun. Math. Phys.} {\bf 249} (2004) 383--415,
  [\href{http://xxx.lanl.gov/abs/hep-th/0306062}{{\tt hep-th/0306062}}].

\bibitem{Mathai:2004qc}
V.~Mathai and J.~M. Rosenberg, {\it On mysteriously missing {T}-duals,
  {$H$}-flux and the {T}-duality group},
  \href{http://xxx.lanl.gov/abs/hep-th/0409073}{{\tt hep-th/0409073}}.

\bibitem{Hull:2004in}
C.~M. Hull, {\it A geometry for non-geometric string backgrounds},  {\em JHEP}
  {\bf 10} (2005) 065, [\href{http://xxx.lanl.gov/abs/hep-th/0406102}{{\tt
  hep-th/0406102}}].

\bibitem{Bouwknegt:2004ap}
P.~Bouwknegt, K.~Hannabuss, and V.~Mathai, {\it Nonassociative tori and
  applications to {T-duality}},  {\em Commun. Math. Phys.} {\bf 264} (2006)
  41--69, [\href{http://xxx.lanl.gov/abs/hep-th/0412092}{{\tt
  hep-th/0412092}}].

\bibitem{Shelton:2005cf}
J.~Shelton, W.~Taylor, and B.~Wecht, {\it Nongeometric flux compactifications},
   {\em JHEP} {\bf 10} (2005) 085,
  [\href{http://xxx.lanl.gov/abs/hep-th/0508133}{{\tt hep-th/0508133}}].

\bibitem{Ellwood:2006my}
I.~Ellwood and A.~Hashimoto, {\it Effective descriptions of branes on
  non-geometric tori},  \href{http://xxx.lanl.gov/abs/hep-th/0607135}{{\tt
  hep-th/0607135}}.

\bibitem{Mathai:2004qq}
V.~Mathai and J.~M. Rosenberg, {\it T-duality for torus bundles via
  noncommutative topology},  {\em Commun. Math. Phys.} {\bf 253} (2004)
  705--721, [\href{http://xxx.lanl.gov/abs/hep-th/0401168}{{\tt
  hep-th/0401168}}].

\bibitem{Mathai:2005fd}
V.~Mathai and J.~Rosenberg, {\it T-duality for torus bundles with {H}-fluxes
  via noncommutative topology. {II}: The high-dimensional case and the
  {T}-duality group},  {\em Adv. Theor. Math. Phys.} {\bf 10} (2006) 123--158,
  [\href{http://xxx.lanl.gov/abs/hep-th/0508084}{{\tt hep-th/0508084}}].

\bibitem{Dabholkar:2005ve}
A.~Dabholkar and C.~Hull, {\it Generalised {T}-duality and non-geometric
  backgrounds},  {\em JHEP} {\bf 05} (2006) 009,
  [\href{http://xxx.lanl.gov/abs/hep-th/0512005}{{\tt hep-th/0512005}}].

\bibitem{Hull:2006va}
C.~M. Hull, {\it Doubled geometry and {T}-folds},
  \href{http://xxx.lanl.gov/abs/hep-th/0605149}{{\tt hep-th/0605149}}.

\bibitem{Hull:2006qs}
C.~M. Hull, {\it Global aspects of {T}-duality, gauged sigma models and {T}-
  folds},  \href{http://xxx.lanl.gov/abs/hep-th/0604178}{{\tt hep-th/0604178}}.

\bibitem{Lawrence:2006ma}
A.~Lawrence, M.~B. Schulz, and B.~Wecht, {\it D-branes in nongeometric
  backgrounds},  \href{http://xxx.lanl.gov/abs/hep-th/0602025}{{\tt
  hep-th/0602025}}.

\bibitem{Hellerman:2006tx}
S.~Hellerman and J.~Walcher, {\it Worldsheet {CFTs} for flat monodrofolds},
  \href{http://xxx.lanl.gov/abs/hep-th/0604191}{{\tt hep-th/0604191}}.

\bibitem{GLW}
M.~Gra$\tilde{{\rm n}}$a, J.~Louis, and D.~Waldram, {\it {\rm{in
  preparation}}}.

\bibitem{Hitchin:2004ut}
N.~Hitchin, {\it Generalized {Calabi--Yau} manifolds},  {\em Quart. J. Math.
  Oxford Ser.} {\bf 54} (2003) 281--308,
  [\href{http://xxx.lanl.gov/abs/math.dg/0209099}{{\tt math.dg/0209099}}].

\bibitem{Gualtieri:2003dx}
M.~Gualtieri, {\it Generalized complex geometry},
  \href{http://xxx.lanl.gov/abs/math.dg/0401221}{{\tt math.dg/0401221}}.

\bibitem{Kapustin:2003sg}
A.~Kapustin, {\it Topological strings on noncommutative manifolds},  {\em Int.
  J. Geom. Meth. Mod. Phys.} {\bf 1} (2004) 49--81,
  [\href{http://xxx.lanl.gov/abs/hep-th/0310057}{{\tt hep-th/0310057}}].

\bibitem{Pestun:2006rj}
V.~Pestun, {\it Topological strings in generalized complex space},
  \href{http://xxx.lanl.gov/abs/hep-th/0603145}{{\tt hep-th/0603145}}.

\bibitem{Cattaneo:1999fm}
A.~S. Cattaneo and G.~Felder, {\it A path integral approach to the {K}ontsevich
  quantization formula},  {\em Commun. Math. Phys.} {\bf 212} (2000) 591--611,
  [\href{http://xxx.lanl.gov/abs/math.qa/9902090}{{\tt math.qa/9902090}}].

\bibitem{DH}
  M.~R.~Douglas and C.~M.~Hull,
  {\it D-branes and the noncommutative torus},
  JHEP {\bf 9802}, 008 (1998),
  \href{http://xxx.lanl.gov/abs/hep-th/9711165}{{\tt hep-th/9711165}}.


\bibitem{CH}
  C.~S.~Chu and P.~M.~Ho,
  {\it Noncommutative open string and D-brane},
  {\em Nucl. Phys.} {\bf B550}, 151 (1999),
  \href{http://xxx.lanl.gov/abs/hep-th/9812219}{{\tt hep-th/9812219}}.


\bibitem{VS}
  V.~Schomerus,
  {\it D-branes and deformation quantization},
  {\em JHEP} {\bf 9906}, 030 (1999),
   \href{http://xxx.lanl.gov/abs/hep-th/9903205}{{\tt hep-th/9903205}}.


\bibitem{Seiberg:1999vs}
N.~Seiberg and E.~Witten, {\it String theory and noncommutative geometry},
  {\em JHEP} {\bf 09} (1999) 032,
  [\href{http://xxx.lanl.gov/abs/hep-th/9908142}{{\tt hep-th/9908142}}].

\bibitem{Anazawa:1999ub}
M.~Anazawa, {\it D0-branes in a {$H$-field} background and noncommutative
  geometry},  {\em Nucl. Phys.} {\bf B569} (2000) 680--692,
  [\href{http://xxx.lanl.gov/abs/hep-th/9905055}{{\tt hep-th/9905055}}].

\bibitem{Lowe:2003qy}
D.~A. Lowe, H.~Nastase, and S.~Ramgoolam, {\it Massive {IIA} string theory and
  matrix theory compactification},  {\em Nucl. Phys.} {\bf B667} (2003) 55--89,
  [\href{http://xxx.lanl.gov/abs/hep-th/0303173}{{\tt hep-th/0303173}}].

\bibitem{Witten:2000cn}
E.~Witten, {\it Overview of {K}-theory applied to strings},  {\em Int. J. Mod.
  Phys.} {\bf A16} (2001) 693--706,
  [\href{http://xxx.lanl.gov/abs/hep-th/0007175}{{\tt hep-th/0007175}}].

\bibitem{Bouwknegt:2000qt}
P.~Bouwknegt and V.~Mathai, {\it D-branes, {B}-fields and twisted {K}-theory},
  {\em JHEP} {\bf 03} (2000) 007,
  [\href{http://xxx.lanl.gov/abs/hep-th/0002023}{{\tt hep-th/0002023}}].

\bibitem{Witten:1985cc}
E.~Witten, {\it Noncommutative geometry and string field theory},  {\em Nucl.
  Phys.} {\bf B268} (1986) 253.

\bibitem{Cavalcanti2005}
G.~Cavalcanti and M.~Gualtieri, {\it Generalized complex structures in
  nilmanifolds},  {\em J. Symplectic Geom.} {\bf 2} (2004) 393--410,
  [\href{http://xxx.lanl.gov/abs/math.DG/0404451}{{\tt math.DG/0404451}}].

\bibitem{GMT}
M.~Gra$\tilde{{\rm n}}$a, R.~Minasian, M.~Petrini and A.~Tomasiello, {\it
  { A scan for new ${\mathcal{N}}=1$} vacua on twisted tori}, 
   \href{http://xxx.lanl.gov/abs/hep-th/0609124}{{\tt hep-th/0609124}}.

\bibitem{Ooguri:1996ck}
H.~Ooguri, Y.~Oz, and Z.~Yin, {\it D-branes on {Calabi--Yau} spaces and their
  mirrors},  {\em Nucl. Phys.} {\bf B477} (1996) 407--430,
  [\href{http://xxx.lanl.gov/abs/hep-th/9606112}{{\tt hep-th/9606112}}].

\bibitem{Kapustin:2001ij}
A.~Kapustin and D.~Orlov, {\it Remarks on {A}-branes, mirror symmetry, and the
  {F}ukaya category},  {\em J. Geom. Phys.} {\bf 48} (2003)
  [\href{http://xxx.lanl.gov/abs/hep-th/0109098}{{\tt hep-th/0109098}}].

\bibitem{Chiantese:2004pe}
S.~Chiantese, {\it Isotropic {A}-branes and the stability condition},  {\em
  JHEP} {\bf 02} (2005) 003,
  [\href{http://xxx.lanl.gov/abs/hep-th/0412181}{{\tt hep-th/0412181}}].

\bibitem{Marino:1999af}
M.~Mari$\tilde{\mathrm{n}}$o, R.~Minasian, G.~W. Moore, and A.~Strominger, {\it
  Nonlinear instantons from supersymmetric $p$-branes},  {\em JHEP} {\bf 01}
  (2000) 005, [\href{http://xxx.lanl.gov/abs/hep-th/9911206}{{\tt
  hep-th/9911206}}].

\bibitem{Kapustin:2004gv}
A.~Kapustin and Y.~Li, {\it Topological sigma-models with {H}-flux and twisted
  generalized complex manifolds},
  \href{http://xxx.lanl.gov/abs/hep-th/0407249}{{\tt hep-th/0407249}}.

\bibitem{Grange:2004ah}
P.~Grange and R.~Minasian, {\it Modified pure spinors and mirror symmetry},
  {\em Nucl. Phys.} {\bf B732} (2006) 366--378,
  [\href{http://xxx.lanl.gov/abs/hep-th/0412086}{{\tt hep-th/0412086}}].

\bibitem{Koerber:2005qi}
P.~Koerber, {\it Stable {D}-branes, calibrations and generalized {Calabi--Yau}
  geometry},  {\em JHEP} {\bf 08} (2005) 099,
  [\href{http://xxx.lanl.gov/abs/hep-th/0506154}{{\tt hep-th/0506154}}].

\bibitem{BB1}
O.~Ben-Bassat and M.~Boyarchenko, {\it Submanifolds of generalized complex
  manifolds},  {\em J. Symplectic Geom.} {\bf 2} (2004), no.~3 309--355.

\bibitem{BB2}
O.~Ben-Bassat, {\it Mirror symmetry and generalized complex manifolds. {I}.
  {T}he transform on vector bundles, spinors, and branes},  {\em J. Geom.
  Phys.} {\bf 56} (2006), no.~4 533--558.

\bibitem{LM1}
L. Martucci and P. Smyth,
  {\it Supersymmetric D-branes and calibrations on general N = 1 backgrounds},
  {\em JHEP} {\bf 0511} (2005) 048,
  [\href{http://xxx.lanl.gov/abs/hep-th/0507099}{{\tt hep-th/0507099}}].
  
\bibitem{LM2}
L.~Martucci,
  {\it D-branes on general N = 1 backgrounds: Superpotentials and D-terms},
  {\em JHEP} {\bf 0606} (2006) 033,
  [\href{http://xxx.lanl.gov/abs/hep-th/0602129}{{\tt hep-th/0602129}}].


\bibitem{Grange:2005nm}
P.~Grange and R.~Minasian, {\it Tachyon condensation and {D}-branes in
  generalized geometries},  {\em Nucl. Phys.} {\bf B741} (2006) 199--214,
  [\href{http://xxx.lanl.gov/abs/hep-th/0512185}{{\tt hep-th/0512185}}].

\bibitem{Minasian:1997mm}
R.~Minasian and G.~W. Moore, {\it K-theory and {Ramond--Ramond} charge},  {\em
  JHEP} {\bf 11} (1997) 002,
  [\href{http://xxx.lanl.gov/abs/hep-th/9710230}{{\tt hep-th/9710230}}].

\bibitem{Witten:1998cd}
E.~Witten, {\it D-branes and {K}-theory},  {\em JHEP} {\bf 12} (1998) 019,
  [\href{http://xxx.lanl.gov/abs/hep-th/9810188}{{\tt hep-th/9810188}}].

\bibitem{Ben-Bassat:2005ne}
O.~Ben-Bassat, J.~Block, and T.~Pantev, {\it Non-commutative tori and
  Fourier-Mukai duality},  \href{http://xxx.lanl.gov/abs/math.ag/0509161}{{\tt
  math.ag/0509161}}.

\bibitem{Witten:1990wb}
E.~Witten, {\it A note on the antibracket formalism},  {\em Mod. Phys. Lett.}
  {\bf A5} (1990) 487.

\bibitem{Ikeda:2006wd}
N.~Ikeda, {\it Deformation of {Batalin--Vilkovisky} structures},
  \href{http://xxx.lanl.gov/abs/math.sg/0604157}{{\tt math.sg/0604157}}.


\bibitem{Schaller:1994es}
P. Schaller  and T. Strobl, {\it Poisson structure induced (topological) field theories},
  {\em Mod. Phys. Lett.} {\bf A9} (1994) 3129--3136,
  [\href{http://xxx.lanl.gov/abs/hep-th/9405110}{{\tt hep-th/9405110}}].


\bibitem{Kontsevich:1997vb}
    M. Kontsevich,
     {\it Deformation quantization of Poisson manifolds, I},
     {\em Lett. Math. Phys.}
     {\bf 66} (2003) 157--216,
[\href{http://xxx.lanl.gov/abs/q-alg/9709040}{{\tt hep-th/9709040}}].

  


\bibitem{Cornalba:1999ah}
L.~Cornalba, {\it D-brane physics and noncommutative {Yang--Mills} theory},
  {\em Adv. Theor. Math. Phys.} {\bf 4} (2000) 271--281,
  [\href{http://xxx.lanl.gov/abs/hep-th/9909081}{{\tt hep-th/9909081}}].

\bibitem{Jurco:2000fb}
B.~{Jur\u{c}o} and P.~Schupp, {\it Noncommutative {Yang--Mills} from
  equivalence of star products},  {\em Eur. Phys. J.} {\bf C14} (2000)
  367--370, [\href{http://xxx.lanl.gov/abs/hep-th/0001032}{{\tt
  hep-th/0001032}}].

\bibitem{Grana:2005sn}
M.~Gra$\tilde{{\rm n}}$a, R.~Minasian, M.~Petrini, and A.~Tomasiello, {\it
  Generalized structures of {N = 1} vacua},  {\em JHEP} {\bf 11} (2005) 020,
  [\href{http://xxx.lanl.gov/abs/hep-th/0505212}{{\tt hep-th/0505212}}].

\bibitem{Grana:2005ny}
M.~Gra$\tilde{{\rm n}}$a, J.~Louis, and D.~Waldram, {\it Hitchin functionals in
  {N} = 2 supergravity},  {\em JHEP} {\bf 01} (2006) 008,
  [\href{http://xxx.lanl.gov/abs/hep-th/0505264}{{\tt hep-th/0505264}}].

\bibitem{Benmachiche:2006df}
I.~Benmachiche and T.~W. Grimm, {\it Generalized {${\mathcal{N}} = 1$}
  orientifold compactifications and the {H}itchin functionals},  {\em Nucl.
  Phys.} {\bf B748} (2006) 200--252,
  [\href{http://xxx.lanl.gov/abs/hep-th/0602241}{{\tt hep-th/0602241}}].

\bibitem{Grana:2005jc}
M.~Gra$\tilde{{\rm n}}$a, {\it Flux compactifications in string theory: A
  comprehensive review},  {\em Phys. Rept.} {\bf 423} (2006) 91--158,
  [\href{http://xxx.lanl.gov/abs/hep-th/0509003}{{\tt hep-th/0509003}}].

\bibitem{Kapustin:2000aa}
A.~Kapustin and D.~Orlov, {\it Vertex algebras, mirror symmetry, and
  {D}-branes: The case of complex tori},  {\em Commun. Math. Phys.} {\bf 233}
  (2003) 79--136, [\href{http://xxx.lanl.gov/abs/hep-th/0010293}{{\tt
  hep-th/0010293}}].

\bibitem{Zucchini:2004ta}
R.~Zucchini, {\it A sigma model field theoretic realization of {Hitchin's}
  generalized complex geometry},  {\em JHEP} {\bf 11} (2004) 045,
  [\href{http://xxx.lanl.gov/abs/hep-th/0409181}{{\tt hep-th/0409181}}].

\bibitem{Zucchini:2005rh}
R.~Zucchini, {\it Generalized complex geometry, generalized branes and the
  {Hitchin} sigma model},  {\em JHEP} {\bf 03} (2005) 022,
  [\href{http://xxx.lanl.gov/abs/hep-th/0501062}{{\tt hep-th/0501062}}].


\bibitem{Howe:2005je}
P.~S.~Howe, U.~Lindstr\"om and V.~Stojevic, {\it Special holonomy sigma models with boundaries},  {\em JHEP} {\bf 0601} (2006) 159,
  [\href{http://xxx.lanl.gov/abs/hep-th/0507035}{{\tt hep-th/0507035}}].



\bibitem{Hassan:1994mq}
S.~F. Hassan, {\it {$O(d,d; {\mathbb{R}})$} deformations of complex structures
  and extended world sheet supersymmetry},  {\em Nucl. Phys.} {\bf B454} (1995)
  86--102, [\href{http://xxx.lanl.gov/abs/hep-th/9408060}{{\tt
  hep-th/9408060}}].

\bibitem{Hassan:1999bv}
S.~F. Hassan, {\it T-duality, space-time spinors and {RR} fields in curved
  backgrounds},  {\em Nucl. Phys.} {\bf B568} (2000) 145--161,
  [\href{http://xxx.lanl.gov/abs/hep-th/9907152}{{\tt hep-th/9907152}}].

\bibitem{Hassan:1999mm}
S.~F. Hassan, {\it {SO(d,d)} transformations of {R}amond{--}{R}amond fields and
  space- time spinors},  {\em Nucl. Phys.} {\bf B583} (2000) 431--453,
  [\href{http://xxx.lanl.gov/abs/hep-th/9912236}{{\tt hep-th/9912236}}].

\bibitem{Lunin:2005jy}
O.~Lunin and J.~M. Maldacena, {\it Deforming field theories with {U(1) $\times$
  U(1)} global symmetry and their gravity duals},  {\em JHEP} {\bf 05} (2005)
  033, [\href{http://xxx.lanl.gov/abs/hep-th/0502086}{{\tt hep-th/0502086}}].

\bibitem{Minasian:2006hv}
R.~Minasian, M.~Petrini, and A.~Zaffaroni, {\it Gravity duals to deformed {SYM}
  theories and generalized complex geometry},
  \href{http://xxx.lanl.gov/abs/hep-th/0606257}{{\tt hep-th/0606257}}.

\bibitem{Strominger:1996it}
A.~Strominger, S.-T. Yau, and E.~Zaslow, {\it Mirror symmetry is {T}-duality},
  {\em Nucl. Phys.} {\bf B479} (1996) 243--259,
  [\href{http://xxx.lanl.gov/abs/hep-th/9606040}{{\tt hep-th/9606040}}].

\bibitem{Grange:2004ra}
P.~Grange, {\it Branes as stable holomorphic line bundles on the non-
  commutative torus},  {\em JHEP} {\bf 10} (2004) 002,
  [\href{http://xxx.lanl.gov/abs/hep-th/0403126}{{\tt hep-th/0403126}}].

\end{thebibliography}

\providecommand{\href}[2]{#2}\begingroup\raggedright\endgroup

\end{document}